\documentclass[eqsecnum,floats,aps,prd,floatfix,titlepage,tightenlines]{revtex4} 

\usepackage{graphicx}
\usepackage{graphics}
\usepackage{bm}
\usepackage{amssymb}
\usepackage{amsmath}
\usepackage{mathrsfs}

\begin{document}

\title{Gravitons and Temperature Fluctuation Correlations from Inflation}

\author{L. H. Ford}
 \email{ford@cosmos.phy.tufts.edu}
 \affiliation{Institute of Cosmology, Department of Physics and Astronomy, Tufts University, Medford, Massachusetts 02155, USA}
     
\author{I-Tai Ho }
 \email{yidaihoap@gmail.com}
 \affiliation{Department of Physics, National Taiwan Normal University, Taipei 116, Taiwan }
     
\author{ Chun-Hsien Wu}
 \email{chunwu@scu.edu.tw}
 \affiliation{Department of Physics, Soochow University, Taipei 111, Taiwan }

\begin{abstract}
Inflationary tensor perturbations are treated as arising from a bath of gravitons produced by quantum particle creation at the end of inflation. 
We calculate the correlation function of the CMB temperature fluctuations produced by these gravitons in a model with an infrared cut off.
The CMB photons are emitted from within a last scattering shell of finite thickness in redshift. We find the correlation function in terms
of the separation of a pair of spacetime points of emission in both angle and redshift. In both variables, there is a significant amount of 
anti-correlation. The anti-correlation minimum has a relative magnitude compared to the central correlation maximum of about $20 \%$
in angle and $50 \%$ in redshift. 
\end{abstract}

\maketitle

\baselineskip=14pt

\section{Introduction}
\label{sec:intro}

This work will re-examine inflationary tensor perturbations from a viewpoint which emphasizes their quantum origin and nature.  
 Starobinsky~\cite{S79} was apparently the first author to suggest that a period of deSitter expansion could produce an observable
 spectrum of gravity wave perturbations in the subsequent universe. This early work was soon followed by contributions from several
 authors, including   Rubakov,  Sazhin, and Veryaskin~\cite{RSV82},  Fabbri and  Pollack~\cite{FP83},  Abbott and Wise~\cite{AW84b},
 Abbott and Harari~\cite{AH86},    Polnarev~\cite{Polnanev}, and Allen~\cite{A88}.

A key result found in Refs.~\cite{FP83,Polnanev} is that a classical gravitational wave propagating in an expanding universe induces an
anisotropic pattern of redshifts on photons also propagating in this universe. These are the tensor perturbations in the Cosmic Microwave
Background (CMB) which may be observable. When these anisotropically redshifted photons scatter from electrons, a particular polarization 
pattern (the B-modes) is induced and may serve as an observational signal of primordial gravity waves. For a review, see  Kamionkowski 
and  Kovetz~\cite{KK16}.

Inflation does not produce a single mode gravitational wave, but rather a spectrum of wave numbers and polarization states. This mixture is often
described as a stochastic bath of gravity waves. However, a deeper approach involves quantum creation of gravitons by the expansion of the universe,
a process first described by Parker~\cite{Parker69} and applied to graviton creation in Refs.~\cite{Grishchuk:1975,FordParker:1977}. For a review, see Ref.~\cite{pp-review}. 
 Calculations of graviton 
creation at the end of inflation were given by Abbott and Harari~\cite{AH86} and by Allen~\cite{A88}. The quantum state of the created gravitons is predicted
to be  highly nonclassical, specifically a squeezed vacuum state~\cite{GS89} .There has been an extensive discussion in the literature as to whether there
is a quantum to classical transition and to whether the gravitons undergo decoherence~\cite{PS96,GS19,Kanno22,Burgess23,Ning23}.

In the present paper, we explore the hypothesis that no quantum to classical transition occurs, and the gravitons remain in a nonclassical state when 
they interact with the CMB photons. The outline of the paper is the following: In Sect.~\ref{sec:creation} we review quantum graviton creation at the end of inflation,
and discuss the resulting quantum state in Sect.~\ref{sec:state}. The effects of the gravitons on the CMB photons is treated in  Sect.~\ref{sec:CMB}.
Section~\ref{sec:Temp} contains our main results on the temperature correlation function and its physical implications. Section~\ref{sec:final} gives a 
summary.

\section{Graviton creation at the end of inflation}
\label{sec:creation}

This section will review the basic principles of quantum particle creation in an expanding universe and application to inflationary cosmology.

\subsection{Gravitons in a Spatially Flat Universe}
\label{sec:flat}

 Here we consider the spatially flat Friedman-Robertson-Walker spacetime with metric
 \begin{equation}
ds^2 = -dt^2 + a^2(t) d{\bf x}^2 = a^2(\eta)\,
       \bigl( - d\eta^2  + d{\bf x}^2\bigr)\,,
       \label{eq:metric}
\end{equation}
where $t$ and $\eta$ are the comoving and conformal time coordinates, respectively. 
We are interested in treating gravitons propagating on this background spacetime. It was shown by Lifshitz\cite{Lifshitz}  that in the transverse, 
tracefree gauge, which eliminates all gauge freedom, each of the two independent polarization states is equivalent to a massless,
minimally coupled scalar field which satisfies
\begin{equation}
\Box \varphi = \nabla_\mu \,\nabla^\mu \varphi =  0 \,. 
\label{eq:KG}
\end{equation}
The positive norm plane wave solutions of this equation may be taken to be 
\begin{equation}
f_{\bf k}({\bf x},\eta)={\frac{\sqrt{16 \pi}\, \ell_P\, e^{i{\bf k\cdot x}}}{a(\eta)\sqrt {(2\pi)^3}}}\,\, \chi_k(\eta), 
\label{eq:f-def}
\end{equation}
where $\ell_P$ is the Planck length and we use the normalization for graviton modes, which is a factor of $\sqrt{16 \, \pi}\, \ell_P$ times that for 
the scalar field, Here $\chi_k(\eta)$ satisfies 
\begin{equation}
{\frac{d^2\chi_k}{d\eta^2}} + \left[ k^2  - \frac{1}{6}\, a^2(\eta)\, R(\eta)\right]\chi_{k} =0 \, ,
  \label{eq:chieq}
\end{equation}
where $R(\eta)$ is the scalar curvature.

First consider the spacetime of the inflationary model, deSitter space, where we may take
\begin{equation}
 a(\eta) = - \frac{1}{H\, \eta}\,,
 \label{eq:inflation-scale}
 \end{equation}
where $H$ is the Hubble parameter and $\eta_0 < \eta < \eta_R < 0$. Here $\eta_0$ denotes the time when inflation began and $\eta_R$ that when it ends
in reheating. Here the scalar curvature is $R = 12 \, H^2$. The solutions of Eq.~\eqref{eq:chieq} may be expressed in terms of Hankel functions 
as
 \begin{equation}
 \chi_k(\eta) = \frac{\eta^{3/2}}{\sqrt{2\, k}} \left[ c_1(k)\, H^{(1)}_{3/2} (k \eta) + c_2(k)\, H^{(2)}_{3/2} (k \eta) \right] \,,
 \end{equation}
where $|c_2(k)|^2 - |c_1(k)|^2 = 1.$
 
 If we set $c_1 = 0$ and $c_2 =1$, the resulting mode function
 \begin{equation}
 \chi^{BD}_k(\eta) = \frac{\eta^{3/2}}{\sqrt{2\, k}} \, H^{(2)}_{3/2} (k \eta) =  \frac{1}{\sqrt{2\, k}} \, \left( 1- \frac{i}{k\, \eta}\right) \;   {\rm e}^{-i \,  k \eta} \,,
 \label{eq:BD}
 \end{equation}
defines the Bunch-Davies vacuum state of deSitter space~\cite{BD78}. In the case of a massive scalar field, it is the unique deSitter invariant state, and is analogous
 to the Minkowski vacuum, the unique Lorentz invariant state in flat spacetime. However,  the massless scalar field has a infrared divergent two-point function,
 $\langle \varphi(x)\, \varphi(x')$, so the Bunch-Davies state is not well defined for either the massless scalar or the graviton fields. There is a class of states 
 which are free of infrared divergences~\cite{FordParker:1977b,FV86}. If we require that $c_1(k) + c_2(k) \rightarrow 0$ as $k \rightarrow 0$, then the two-point function is infrared finite.
 This class may be said to be Bunch-Davies-like if we also have $c_1(k) \approx 0$ and $|c_2(k)| \approx 1$ for $k \agt k_c$ for some small cutoff scale $k_c$.
 We can view this scale as set by some initial condition before the onset of inflation.
 If $k_c$ is sufficiently small that the associated modes have not entered the horizon since the end of inflation, then they seem to play no role in present 
 cosmological observations.  In this case, we may take Eq.~\eqref{eq:BD} as the initial graviton mode before reheating.

\subsection{The transition from Inflation to a Radiation Dominated Universe}
\label{sec:transition}

The creation of gravitons at the end of inflation may be viewed as gravitational particle creation described by a Bogolubov transformation. Here we take the 
in-modes to be the Bunch-Davies-like modes given by Eqs.~\eqref{eq:f-def} and \eqref{eq:BD}.  In the subsequent radiation dominated phase, the
scalar curvature vanishes, $R=0$, so the positive unit norm mode functions become
 \begin{equation}
F_{\bf k}({\bf x},\eta)=\frac{ \sqrt{16 \pi}\, \ \ell_P\,  e^{i({\bf k\cdot x} - k\, \eta)}  }{\sqrt{2k \,(2\pi)^3}\,  a(\eta) }\,.
\label{eq:F-def}
\end{equation}
 We are interested in very long wavelength gravitons modes which are far outside the horizon at the end of inflation, so $k \ll H$. These modes are not 
 sensitive to the finite duration of reheating, so we assume a sharp transition in which the scale factor is given by Eq.~\eqref{eq:inflation-scale} for 
 $\eta < -H^{-1}$ and by
 \begin{equation}
 a(\eta) = H\, \eta + 2
 \label{eq:rad-dom-scale}
 \end{equation}
for  $\eta > -H^{-1}$. Note that we take the reheating time to be $\eta_R = -H^{-1}$ and have set $a(\eta_R) = 1$.

The mode functions in the in and out regions are related by a Bogolubov transformation of the form
 \begin{equation}
 f_{\bf k}({\bf x},\eta)= \alpha^*_k\,  F_{\bf k}({\bf x},\eta) + \beta^*_k \, F^*_{\bf k}({\bf x},\eta) \,.
 \label{eq:fF}
 \end{equation}
 where
  \begin{equation}
 |\alpha_k|^2 -  |\beta_k|^2 = 1\,,
 \label{eq:alphabeta2}
 \end{equation}
 The Bogolubov coefficients may be found by using Eq.~\eqref{eq:fF} and its first time derivative at $\eta = \eta_R$, using the fact that both $a(\eta)$
 and $a'(\eta)$ are continuous at this point. The results~\cite{AH86,A88}  are
 \begin{equation}
 |\alpha_k|^2 = 1+ \frac{H^4}{4\, k^4} \, ,
 \label{eq:alpha}
 \end{equation}
 and 
 \begin{equation}
 |\beta_k|^2 =  \frac{H^4}{4\, k^4} \, .
 \label{eq:beta}
 \end{equation}
 
 The mean number of gravitons created in a given mode is  $\langle N_k \rangle =  |\beta_k|^2$. Here that number is very large. The energy spectrum of
 created gravitons is scale invariant in the sense that the graviton energy density in wave number interval $d^3 k$ is proportional to 
 $k\,  |\beta_k|^2 \, d^3 k \propto dk/k$, which independent of the choice of scale for $k$. Note that in a spatially flat expanding universe, linear three-momentum 
 is conserved, so the gravitons are created in pairs with equal and opposite wave vectors, ${\bf k}$ and ${- \bf k}$ 
 
 \section{The Quantum State of the Created Gravitons}
 \label{sec:state}
 
 \subsection{Representation in the Out-Fock Space}
 
 The quantum particle creation process is being described in the Heisenberg picture, so the quantum state of the system, $ |\psi_0 \rangle$, 
 does not change in time. However, its description is very different in the deSitter space in-region from that in the radiation dominated out-region.
 In the in-region, it is one of several infrared finite choices of vacuum state. In the out-region, it is a highly excited state with very large numbers of 
 particles in modes for which $k \ll H$.
 
The quantum state of the created particles in the out-Fock space  may be shown to be
 \begin{equation}
 |\psi_0 \rangle= c_0 \sum_{\{n_{\bf k}\}}  \prod_{\bf k}  \gamma_k^{n_{\bf k}}   \; |\{n_{\bf k}\} \rangle \,,
 \label{eq:state}
 \end{equation}
 where $\gamma_k = \beta_k/\alpha_k$.
Here $\{n_{\bf k}\}$ denote a set of occupation numbers; for each mode ${\bf k}$ there is non-negative integer $n_{\bf k}$ which gives the number of particles in mode 
${\bf k}$
and the number of particles in mode ${-\bf k}$. In our enumeration, we may require $k_z \geq 0$ to avoid over counting. The overall constant $c_0$ satisfies
 \begin{equation}
 |c_0| = \left(\prod_{\bf k} |\alpha_k| \right)^{-1} =  \prod_{\bf k} \sqrt{1- |\gamma_k|^2} \,,
 \end{equation}
 where the latter form follows from Eq.~\eqref{eq:alphabeta2}.
This quantum state is  a multi-mode squeezed vacuum state, containing all possible numbers of pairs of correlated particles. Strictly, Eq.~\eqref{eq:state}
applies to the massless scalar field. For gravitons, we need to include the polarization among the mode labels. This will be explicitly addressed in
Sect~\ref{sec:redshifts},  where a sum over polarizations will be performed.

In the single mode case, only one type of pair is present. Here we have
\begin{equation}
 |\psi_0 \rangle_1 =   \sqrt{1- |\gamma_k|^2}\, \sum_{n=0}^\infty \gamma_k^n   \; | n,n \rangle \,,
 \label{eq:state-1}
 \end{equation}
where $| n,n \rangle$ denotes a state containing $n$ pairs, with one member of the pair in mode ${\bf k}$, and the other in mode ${-\bf k}$.

The expectation value of the field operator vanishes in this state,
 \begin{equation}
 \langle \psi_0| \varphi(x) |\psi_0 \rangle = 0,
 \end{equation}
  due to the fact that
$ \langle m,m| a_{\bf k} |n,n \rangle \propto  \langle m,m | n-1,n \rangle = 0$ for all integers $m$ and $n$. This holds for both the one-mode case of
Eq.~\eqref{eq:state-1}, and the multi-mode case of Eq.~\eqref{eq:state}. As a result, the state $ |\psi_0 \rangle_1$ does not  describe a classical wave.

\subsection{The Observable Part of the State}
\label{sec:observable}

 It is well known in quantum theory that outcomes of 
observations can depend crucially upon the details of the question which is addressed, or the specific measurement which is undertaken.
At least in the cosmological case, we can only observe one member of a pair of correlated gravitons. This can be seen if we adopt a set of wave packet mode
 functions, rather than plane waves with a precisely defined wave vector. Consider a set of wave packets with a finite but small bandwidth, $\Delta k \ll k = |{\bf k}|$,
 where ${\bf k}$ is the peak wave vector of a packet. These packets will have a spatial extent of order $\Delta x \approx 1/\Delta k$.  

\begin{figure}[htbp]
\includegraphics[scale=0.2]{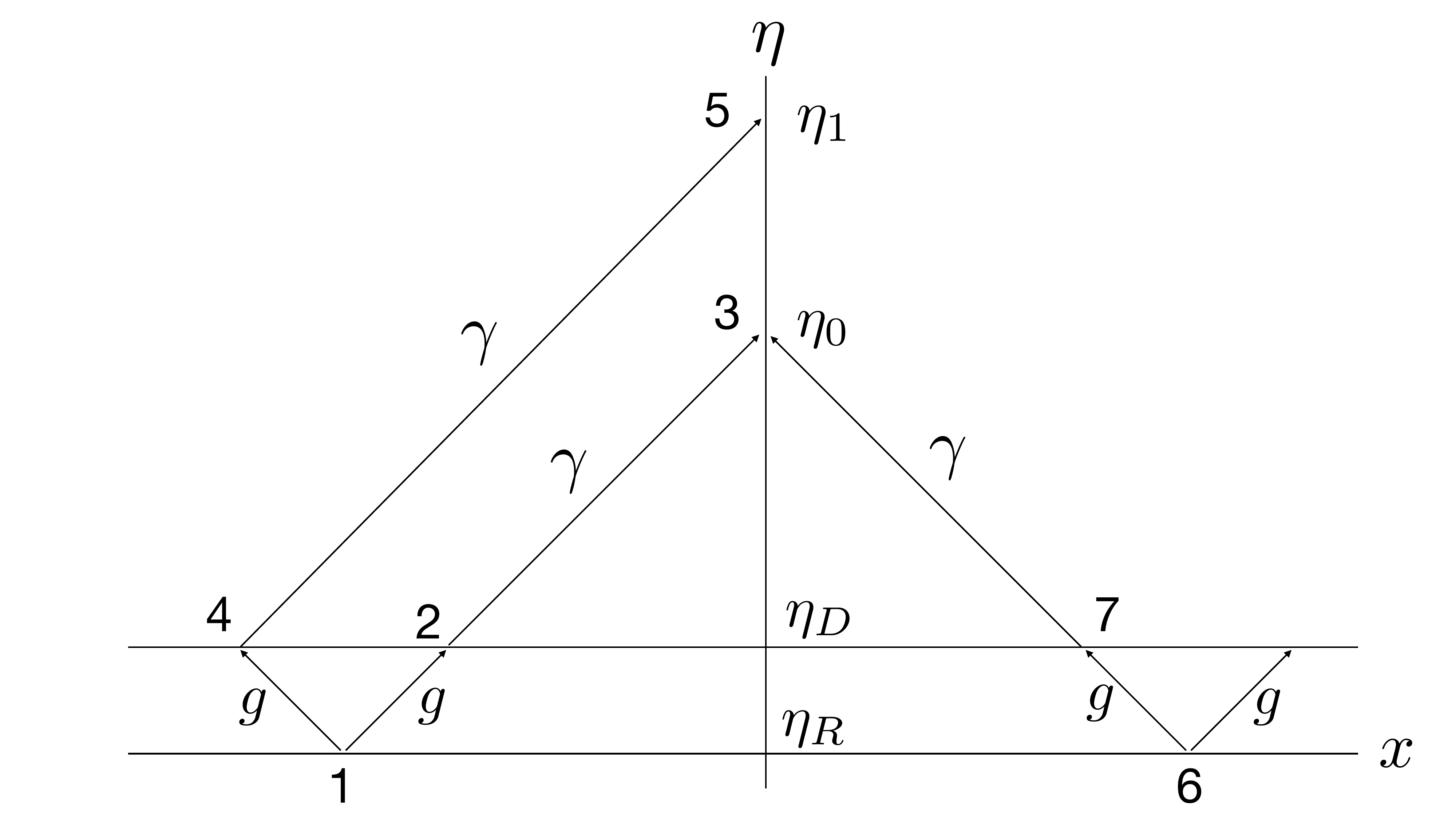}
\caption{A spacetime diagram illustrating the history of graviton ($g$) and photon  ($\gamma$) wavepackets which may produce observable features in the CMB. 
Here $\eta_R$,
$\eta_D$, and $\eta_0$ are the conformal times associated with reheating, decoupling, and the present, respectively. At event $1$, entangled gravitons are created,
with one packet traveling to event $2$, where it influences a photon packet, which reaches Earth at $3$. The other entangled graviton mode goes in the opposite
 direction, and reaches the last scattering region at $\eta = \eta_D$ at event $4$. There it influences photons which reach Earth at event $5$, when $\eta = \eta_1$,
 about $10^9$ years after event $3$. 
 At the same time that the photons from $2$ are observed, there are photons arriving from the opposite direction which were emitted at event $7$ and were 
 influenced by gravitons created at $6$.}
\label{fig:STdiagram}
\end{figure}

In the wave packet basis, the correlated pairs of gravitons are created into a pair of packets peaked about values of ${\bf k}$ with equal magnitude, 
but opposite sign.  This means that the two packets leave the region of their creation in opposite directions. A given observer will be able to observe 
at most one of these gravitons. This is illustrated in Fig,~\ref{fig:STdiagram}   for the case of gravitons which influence the redshifts of CMB photons.

Suppose we are making measurements of local field operators in a spacetime region ${\cal R}$, and using a complete set of wave packets with mode labels
$\{j\}$. It is convenient to divide the set $\{j\}$ into two subsets, $\{j_1\}$ and $\{j_2\}$, where the $j_1$ modes are nonzero in  ${\cal R}$ , and the $j_2$ modes 
vanish there and do not contribute to local expectation values in ${\cal R}$. Now we can rewrite Eq.~\eqref{eq:state} as
 \begin{equation}
 |\psi_0 \rangle= c_0 \sum_{\{n_j \}}  \prod_j \gamma_k^{n_j}   \; |\{n_{j_1} \}, \{n_{j_2} \},\rangle \,.
 \label{eq:state12}
 \end{equation}
Equation~\eqref{eq:state-1} is the special case where there is one $j_1$ mode and one $j_2$ mode.

If $i$ and $\ell$ are two modes in set  $j_1$, then the expectation values in state $ |\psi_0 \rangle$ of the associated creation and annihilation 
operators are
\begin{equation}
 \langle a_i \rangle = \langle a_i^\dagger\rangle = \langle a_i  a_\ell \rangle = \langle a^\dagger_i a^\dagger_\ell \rangle = 0 \,, 
\label{eq:a-rel1}
 \end{equation}
but
\begin{equation}
  \langle a^\dagger_i a_\ell \rangle = \delta_{i\ell}\, |\beta_i|^2 \,.
   \label{eq:a-rel2}
 \end{equation}
This implies that expectation values of linear field operators vanish in this state, but those of quadratic operators can be nonzero.
Furthermore, the effects of the state $ |\psi_0 \rangle$  in ${\cal R}$ are equivalent to a density matrix.

Let $a$ and $a^\dagger$ be associated with any given mode in set  $j_1$, and define a coordinate by
\begin{equation}
 x = \frac{1}{\sqrt{2}} (a + a^\dagger)
 \end{equation}
and a conjugate momentum by
\begin{equation}
 p = \frac{i}{\sqrt{2}} (a - a^\dagger) \,.
 \end{equation}
Then  $\langle x \rangle =  \langle p \rangle = 0 $ \,, but the ubcertainties are nonzero: $\Delta x = \sqrt{x^2} = \Delta p$. Furthermore
the uncertainty relation becomes
 \begin{equation}
 \Delta x \; \Delta p = |\beta|^2 + \frac{1}{2} \,.
 \end{equation}
In general, $ |\psi_0 \rangle$ is not a state of minimum uncertainty, unlike the Stoller squeezed states, and will have very large uncertainty
when $|\beta_j| \gg1$. This implies large fluctuations, as evidenced the large variance of linear field operators.

\subsection{An Electromagnetic Analog Model}
\label{sec:EM-analog}

\begin{figure}[htbp]
\includegraphics[scale=0.2]{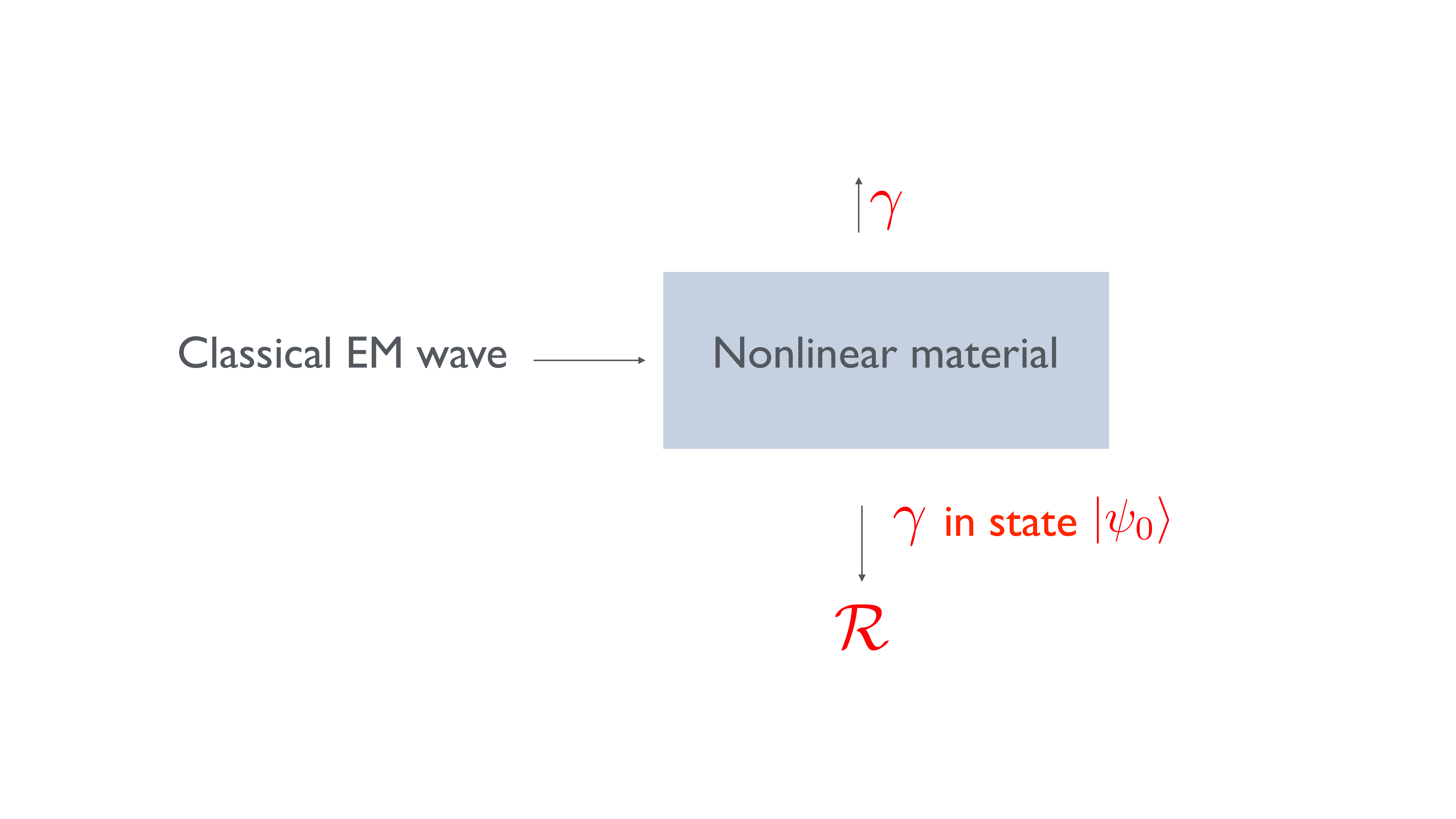}
\caption{ The quantum creation of photons in a squeezed state in a crystal of nonlinear material is illustrated. A strong classical electromagnetic wave
enters the crystal on the left. Correlated pairs of photons are emitted from the crystal in opposite directions. Experiments performed  in regions ${\cal R}$
are sensitive to only one member of each pair.}
\label{fig:EM-analogy}

\end{figure}

There is a simple analog involving a squeezed state of photons which is illustrated in Fig.~\ref{fig:EM-analogy}. A strong classical electromagnetic wave
in a nonlinear material with a nonzero second or third order susceptibility creates a time dependent effective index of refraction for vacuum
 electromagnetic modes. This leads to mixing of positive and negative frequencies and quantum creation of photons, an effect analogous to the
 quantum creation of gravitons in inflation discussed in Sect.~\ref{sec:creation}.  The photons are emitted in correlated pairs in opposite directions
 from the material in a quantum state of the form of $|\psi_0 \rangle$ given in Eq.~\eqref{eq:state12}.

\subsection{Scalar Correlation Functions}
\label{sec:corr}

In this subsection, we focus on the case of a single massless, minimally coupled scalar field $\varphi(x)$, which is equivalent to one polarization state
of the graviton field in the transverse tracefree  gauge. Assume that we are in the radiation dominated era, when the mode functions may be
given by Eq.~\eqref{eq:F-def} and the field operator may be expressed as
 \begin{equation}
\varphi(x)  = \sum_{\bf k} \left[  a_{\bf k}  F_{\bf k}(x) + a^\dagger_{\bf k}  F^*_{\bf k}(x) \right] \,,
 \end{equation}
where the $a_{\bf k}$ satisfy the relations in Eqs.~\eqref{eq:a-rel1} and \eqref{eq:a-rel2} in a state of the form of  $ |\psi_0 \rangle$.  Next define 
a correlation operator by
 \begin{equation}
 \kappa(x,x') = \varphi(x)\, \varphi(x')
 \end{equation}
 and a correlation function by
 \begin{equation}
 C(x,x') = \langle   \kappa(x,x')    \rangle\,.
 \end{equation}
Recall that $\langle   \varphi(x)  \rangle = 0$, so $C(x,x)$ is the variance of the fluctuations of   $\varphi(x)$. If $|\beta_j|^2 \gg 1$, then
\begin{equation}
 C(x,x') \approx  2\,   \sum_{j}  |\beta_j|^2\,  \rm Re,\left[   F^*_j (x')  \, F_j(x)  \right]\,,
\label{eq:corr}
 \end{equation}
where we ignore a term of order  $|\beta_j|^0 $.  

\begin{figure}[htbp]
\includegraphics[scale=0.2]{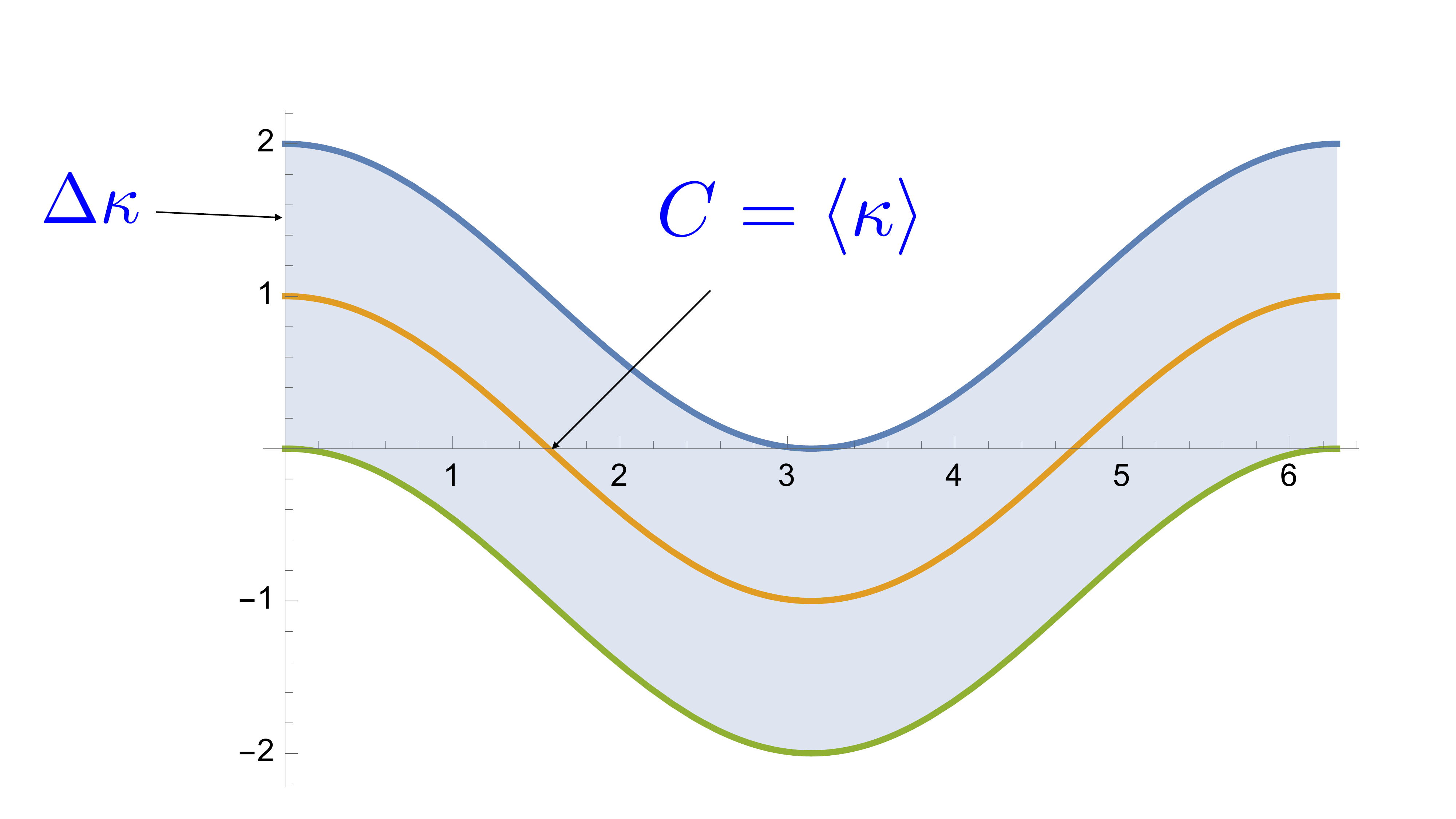}
\caption{ The correlation fluctuations are illustrated. The mean value of the correlation operator $\kappa$, which is the correlation function, undergoes
periodic oscillations. However, the uncertainty, $\Delta \kappa$, is of the same order as the mean value.  }
\label{fig:corr-flucts}
\end{figure}

Consider the case where a single mode $j$ gives the dominant contribution to  $C(x,x')$, so that  $C(x,x') \propto F_j(x)  $, and the correlation function
has the form of a classical wave as a function of $x$ for fixed $x'$. This is similar to the case of a coherent state. However, unlike a highly excited
coherent state, here the fluctuations of  the correlation operator are very large:
\begin{equation}
  \langle   \kappa^2(x,x')    \rangle  = 2 \langle   \kappa(x,x')    \rangle^2 + 4 |\beta_j|^4 \, |F_j (x')  \, F_j(x)|^2 + O(|\beta_j|^2)\,,
 \end{equation}
so the fractional variance of $K$ is of order one
\begin{equation}
  \frac{ \langle   \kappa^2(x,x')    \rangle  -  \langle   \kappa(x,x')    \rangle^2}{\langle   \kappa(x,x')    \rangle^2} = O(|\beta_j|^0)\,.
 \end{equation}
The fluctuations are illustrated in Fig.~\ref{fig:corr-flucts}. Because of the large fractional fluctuations, the correlation effects can only be observed
in an ensemble average.

\section{Effects of Gravity Waves and of Gravitons on CMB Photons}
\label{sec:CMB}

\subsection{Differential Redshifts due to a Classical Gravity Wave}
 \label{sec:redshifts}

 \subsubsection{$+$ Polarization}
 
 The metric may be written as
  \begin{equation}
 ds^2 = a^2(\eta)\, [- d\eta^2 + (1+ h_+)\, dx^2 + (1- h_+)\, dy^2  + dz^2]\,,
 \label{eq:+metric}
 \end{equation}
 where $h_+ = h_+(\eta -z)$. 
 The equation for the four-velocity $u^\mu$ of a timelike geodesic is
 \begin{equation}
 \frac{d u^\mu}{d\tau} = - \Gamma^\mu_{\alpha \beta} \,u^\alpha u^\beta \,,
 \end{equation}
 where $\tau$ is the observer's proper time, and $d\tau = dt = a\, d\eta$.
 The connection coefficient which we need is, to first order in $h_+$,
 \begin{equation}
 \Gamma^\eta_{\eta \eta} = \frac{a'}{a}\,.
 \end{equation}
 Hence the geodesic equation  has the solution $u^\mu = (a^{-1},0,0,0)$.  This tells us that the comoving observers are at rest in the coordinates of
Eq.~\eqref{eq:+metric}. 
Let  $q^\mu$ be the 4-dimensional wave vector of a photon. The angular frequency of this  photon in the comoving observer's frame is
 \begin{equation}
 \omega = - q^\mu \, u_\mu = a(\eta)\, q^\eta \,.
 \label{eq:omega}
 \end{equation}

We may find the wave vector $q^\mu$ either from geodesic equation, or the null vector condition, $q^\mu \, q_\mu = 0$. Here we use the latter. 
In the metric of Eq.~\eqref{eq:+metric} we find
 \begin{equation}
 q^\eta = \sqrt{(1+ h_+) \, (q^x)^2 + (1- h_+) \, (q^y)^2 + (q^z)^2} \,.
 \label{eq:q-eta}
 \end{equation}
In the isotropic case where $h_+ = 0$, we have $q^\mu = q_0^\mu$, where  
 \begin{equation}
  q_0^\mu = (q_0^\eta,q_0^x,q_0^y,q_0^z) =  \omega_0\, a^{-2}(\eta) \, (1, \sin \theta \cos \phi, \sin \theta \sin \phi, \cos \theta)\,.
 \end{equation}
 Here $\theta$ and $\phi$ are the direction angles for the photon in a frame where the gravity wave propagates in the $+z$-direction,
 and $\omega_0$ is the photon angular frequency when $a = 1$
 The factor of $a^{-2}(\eta)$ arises because $q^\mu = \omega_0 \, dx^\mu/d \lambda$  and $\lambda$ is an affine parameter. The latter may be taken
to be given by $d\lambda = a^2\, d\eta$. (See, for example, \cite{Visser23}.)

To first order in $h_+ $, we may place the above expressions for the zero order spatial components $q_0^\mu$ in the right hand side of
Eq.~\eqref{eq:q-eta},  Taylor expand, and use  Eq.~\eqref{eq:omega}  to find
\begin{equation}
 \omega_+ \approx \frac{\omega_0}{a(\eta)} \, \left( 1 + \frac{1}{2} h_+\, \sin^2 \theta\, \cos 2\phi \right)\,.
 \label{eq:omega+}
 \end{equation}
 The pre-factor, ${\omega_0}/{a(\eta)}$, is the redshifted frequency in an isotropic universe, and
the term proportional to $h_+ $ gives the fractional anisotropic redshift due to the gravity wave. Specifically, we define the fractional redshift by
 \begin{equation}
 \Delta  \omega_+ = \frac{ \omega_+  -  \omega_0/a(\eta)}{\omega_0/a(\eta)} =  \frac{1}{2} h_+\, \sin^2 \theta\, \cos 2\phi \,.
 \end{equation}
Apart from a sign difference, this agrees with the
result of Polnarev~\cite{Polnanev}  which is quoted in Ref.~\cite{KK16}

  \subsubsection{$\times$ Polarization}
  
  Now the metric may be written as
  \begin{equation}
 ds^2 = a^2(\eta)\, [- d\eta^2 +  dx^2 + 2 h_\times dx \,dy + dy^2 +  dz^2]\,,
 \label{eq:Xmetric}
 \end{equation}
 where $h_\times = h_\times(\eta -z)$. If we repeat the procedure in the previous subsection, the result is
\begin{equation}
 \omega_\times \approx \frac{\omega_0}{a(\eta)} \, \left( 1 + \frac{1}{2} h_\times\, \sin^2 \theta\, \sin 2\phi \right)\,.
 \label{eq:omegaX}
 \end{equation}
Note that if we let $\phi \rightarrow \phi + \pi/4$, then $\omega_+  \rightarrow  \omega_\times$, as a rotation by $\pi/4$ interchanges the 
two polarizations.
Now the fractional redshift becomes
 \begin{equation}
 \Delta  \omega_\times =  \frac{1}{2} h_\times\, \sin^2 \theta\, \sin 2\phi \,.
 \end{equation}
 
 Note that if $ h_+^2 = h_\times^2 = h^2$, then
 \begin{equation}
 ( \Delta  \omega_+)^2 +  ( \Delta  \omega_\times)^2  =  \frac{1}{4} h^2 \, \sin^4 \theta\,.
 \label{eq:summed-redshift}
 \end{equation}
Thus the dependence upon the angle $\phi$ disappears from a sum over polarization states.

\subsection{The Quantized Graviton Field}
\label{sec:grav-field}

The graviton field operator may be expanded in terms of mode functions as
 \begin{equation}
 h^{\mu\nu}({\bf x},\eta) =  \sum_{{\bf k},\lambda} \left[ {\rm e}^{\mu\nu}({\bf k},\lambda) \, F_{\bf k}({\bf x},\eta) \, a_{{\bf k},\lambda}  +H.c. \right]\, ,
 \end{equation}
where $ {\rm e}^{\mu\nu}({\bf k},\lambda)$ is the polarization tensor for mode ${\bf k},\lambda$. The expectation value of $ h^{\mu\nu}$ in a coherent state
for mode ${{\bf k},\lambda}$ is a classical gravitational wave. Here we use a linear polarization basis, so $\lambda = +, \times$, and the polarization
tensors are implicitly defined by Eqs.~\eqref{eq:+metric} and \eqref{eq:Xmetric}.  

The graviton correlation function may be expressed as
 \begin{equation}
\langle   h^{\mu\nu}({\bf x},\eta) \,  h^{\rho\sigma}({\bf x'},\eta')  = \frac{\ell_P^2}{\pi^2\, a(\eta)\, a(\eta')}\; \int \frac{d^3 k}{k} \, |\beta_k|^2 \,   \sum_{\lambda}  {\rm e}^{\mu\nu} {\rm e}^{\rho\sigma}\; \cos[({\bf k}\cdot({\bf x} - {\bf x'}) -k\, (\eta - \eta')]\,.
\label{eq:grav-corr}
 \end{equation}
  Here we have used Eqs.~\eqref{eq:F-def} and  \eqref{eq:a-rel2}, as well as the convention that $ \sum_{\bf k} \rightarrow \int d^3 k$ in the continuum limit.

\section{The Temperature Correlation Function}
\label{sec:Temp}

The graviton correlation function, Eq.~\eqref{eq:grav-corr}  may be used to compute a CMB temperature fluctuation correlation function using the results of Sect.~\ref{sec:redshifts}.   
Equation~\eqref{eq:summed-redshift} was derived for classical gravity waves, but also applies expectation values in a graviton bath if we replace $h^2$
by an expectation value of a quadratic graviton operator. Specifically, we can convert Eq.~\eqref{eq:grav-corr} into a result for the fractional temperature
fluctuation correlation function:
 \begin{equation}
 C({\bf r},\eta,\eta') =  \frac{\ell_P^2}{4\,\pi^2\, a(\eta)\, a(\eta')}\; \int \frac{d^3 k}{k} \, \sin^4\chi \, |\beta_k|^2 \, \cos[({\bf k}\cdot{\bf r}  -k\, (\eta - \eta')]\,,
\label{eq:temp-corr}
 \end{equation}
 where ${\bf r} =  {\bf x} - {\bf x'}$, and we are interested in the correlation between photons emitted at spacetime point $(\eta,{\bf x})$ with those emitted
 at  $(\eta',{\bf x'})$ .
Here $\chi$  denotes the angle between our line of sight and the graviton wave vector ${\bf k}$. We have used Eq.~\eqref{eq:summed-redshift}
to make the replacement $ \sum_{\lambda}  {\rm e}^{\mu\nu} {\rm e}^{\rho\sigma} \rightarrow  \frac{1}{4}  \, \sin^4 \chi $.

\subsection{The infrared Cutoff $k_0$}
\label{sec:k0}

If we use Eq.~\eqref{eq:beta}, with $ |\beta_k|^2 \propto k^{-4}$, then the integral in Eq.~\eqref{eq:temp-corr} will diverge at the lower limit. This issue is most
likely distinct from the infrared divergence discussed in Sect.~\ref{sec:flat} which required a cutoff at $k = k_e$. The latter was required because gravitons
and massless minimally coupled scalar fields necessarily break deSitter symmetry. The cutoff at $k = k_e$ arises from a memory of pre-inflationary conditions
which is was not erased by a finite amount of inflation. The infrared divergence  carried by $ |\beta_k|^2$ is most likely due to a breakdown of the particle interpretation
of the out-Fock space.  Recall that $ |\beta_k|^2$ is interpreted as the mean number of gravitons in mode $k$ in the out-region. However, it is not clear that graviton 
number is meaningful for modes whose wavelength is far greater than the horizon size. We propose to modify Eq.~\eqref{eq:temp-corr} by restricting the integration to modes
with $k > k_0$, where $k_0$ is the wave number associated with the horizon size at the time when the gravitons interact with the CMB photons. Thus,
Eq.~\eqref{eq:temp-corr} is replaced by
  \begin{equation}
 C({\bf r},\eta,\eta') =  \frac{H^4\, \ell_P^2}{16\,\pi^2\, a(\eta)\, a(\eta')}\; \int^\infty_{k_0} \frac{d k}{k^3} \, \int d\Omega_k \, \sin^4\chi \,  \cos[({\bf k}\cdot{\bf r}  -k\, (\eta - \eta')]\,.
\label{eq:temp-corr2}
 \end{equation}
  
\begin{figure}[htbp]
\includegraphics[scale=0.2]{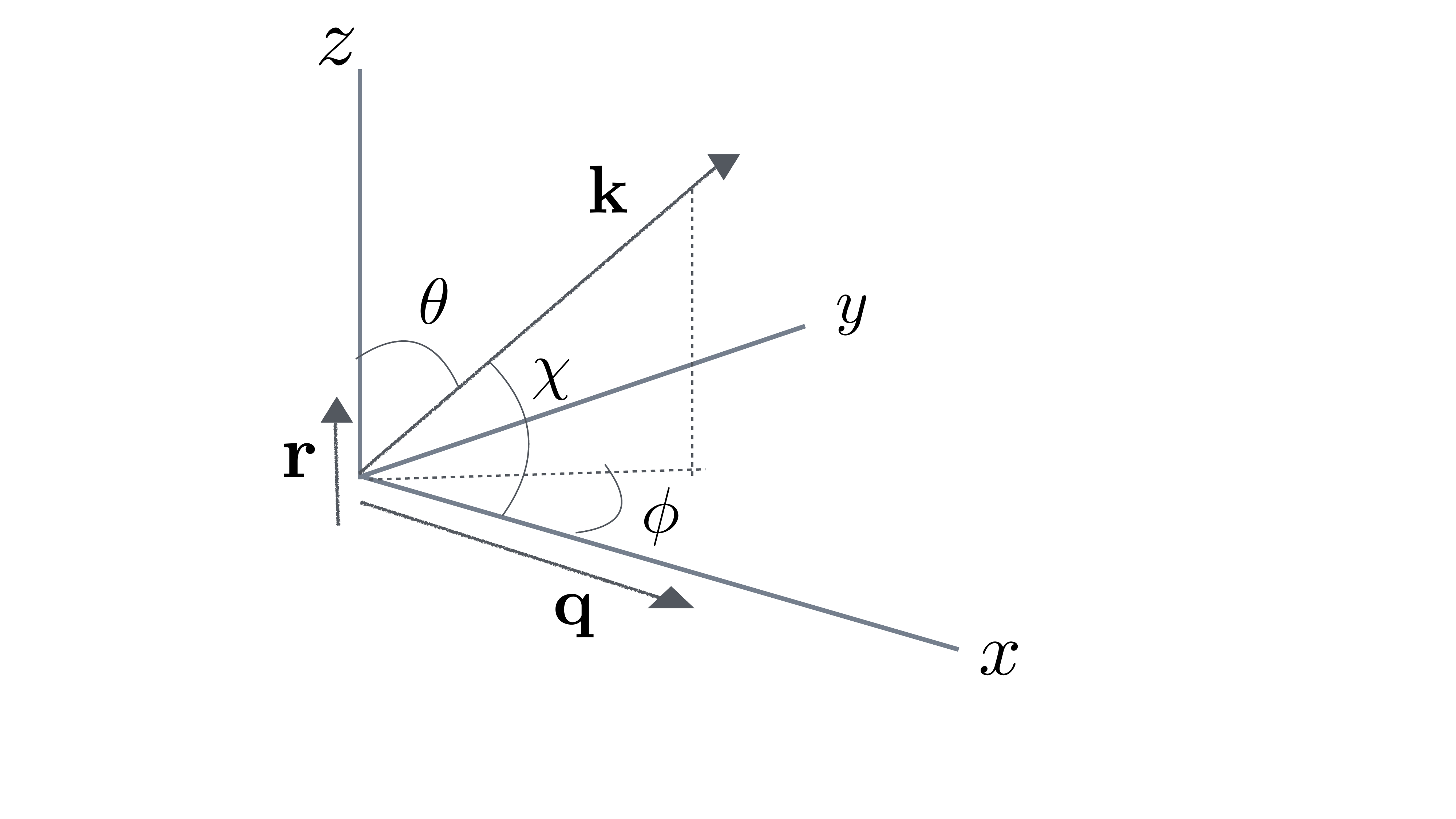}
\caption{ The various vectors appearing in  Eq.~\eqref{eq:temp-corr2} are illustrated in a particular coordinate system. The separation vector ${\mathbf r}$ between the 
two space points lies along the $z$-axis. The photon wave vector  ${\mathbf q}$ lies along the $x$-axis, which is the direction from the point of emission of the photon
to the observer. The graviton wave vector  ${\mathbf k}$ lies along a direction given by the polar angle $\theta$ and the azimuthal angle $\phi$.}
\label{fig:Vectors}
\end{figure}

  The angles involved in the $d\Omega_k$ integration are illustrated in Fig.~\ref{fig:Vectors}. First note that $\cos \chi = \hat{\bf k} \cdot  \hat{\bf q}= \cos \theta \,\cos \phi$.
  Thus $\sin^2 \chi = 1 -(1 -c^2) \, \cos^2 \phi$, with $c = \cos \theta$ and we have
 \begin{equation}
 \int_0^{2 \pi} \sin^4 \chi \, d\phi = \frac{\pi}{4} \, (3 +2 c^2 +3 c^4).
\end{equation}

Now we can write
 \begin{equation}
 C({\bf r},\eta,\eta') =  \frac{H^4\, \ell_P^2}{15\,\pi\, a(\eta)\, a(\eta')\, k_0^2}\; K(\Delta \eta,r)\, ,
\label{eq:temp-corr3}
 \end{equation}
 where we define a reduced correlation function by
  \begin{equation}
  K(\Delta \eta,r) = \frac{15}{64}\,  \int_{-1}^1 dc \, (3 +2 c^2 +3 c^4) \int_1^\infty \frac{du}{u^3} \, \cos[k_0 (c\,  r - \Delta \eta)\, u]\,,
  \label{eq:K-def}
 \end{equation}
 which is normalized by $K(0,0) = 1$.
 The reduced correlation function may be expressed in terms of 
the functions $C_n$ and $S_n$ are defined by
  \begin{equation}
 C_n(\alpha) = \int_1^\infty \frac{du}{u^n} \, \cos(\alpha \, u) \,,
  \label{eq:Cn}
 \end{equation}
 and
  \begin{equation}
 S_n(\alpha) = \int_1^\infty \frac{du}{u^n} \, \sin(\alpha \, u) \,.
  \label{eq:Sn}
 \end{equation}
These functions may be expressed in terms of trigonometric and cosine-integral functions and are given in the Appendix.

\subsection{Results: Case ${\bf r} = 0$ }
  \label{sec:results}

\begin{figure}[htbp]
\includegraphics[scale=0.2]{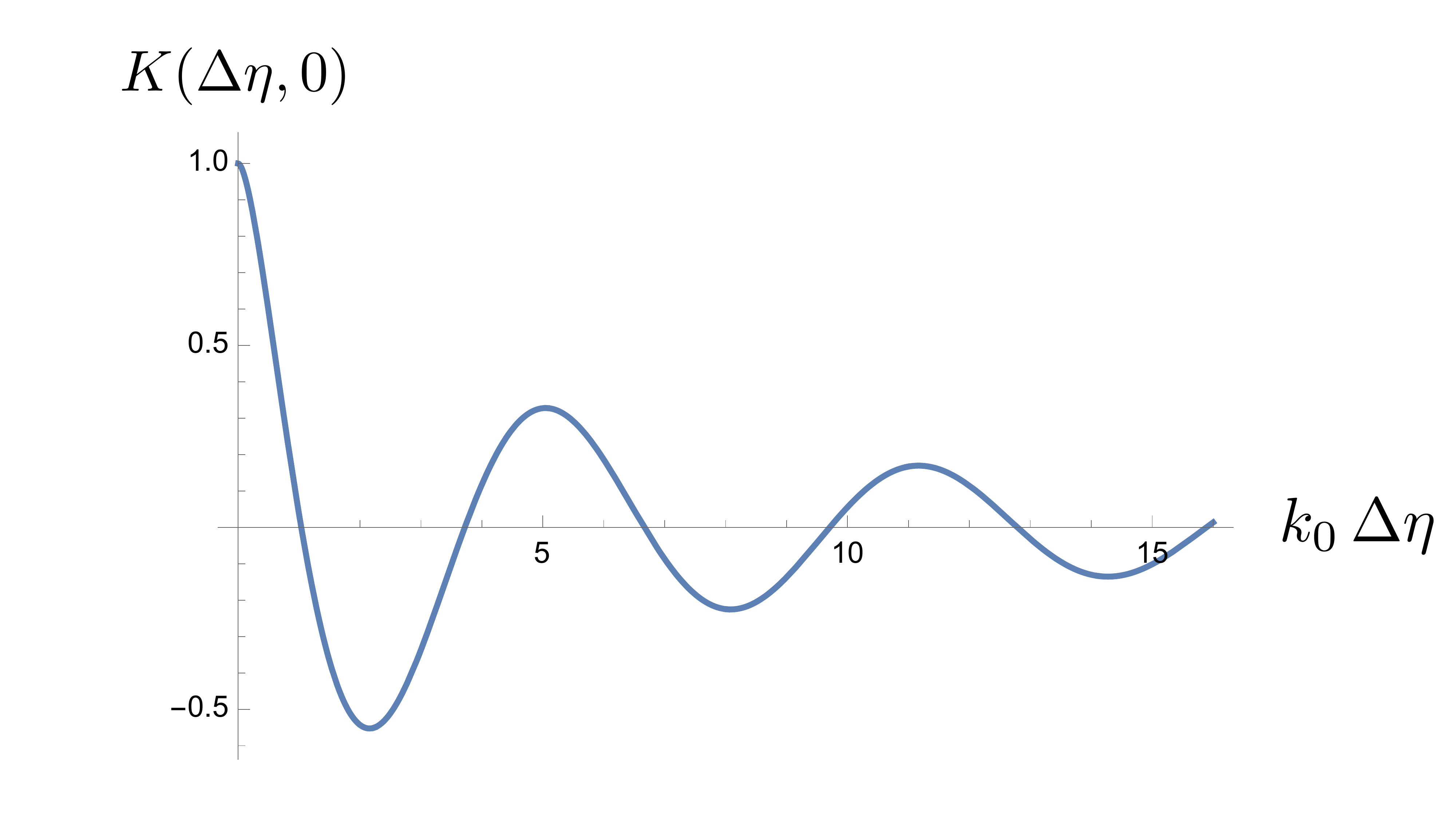}
\caption{ The reduced correlation function, $K(k_0\, \Delta \eta,0)$, for coincident spatial points is plotted. This gives the temperature correlations,
without the effects of the scale factor changes, between regions at different redshifts, but the same spatial location. Note that regions separated by
$k_0\, \Delta \eta \approx 2$ are strongly anti-correlated.}
\label{fig:Corr-plot0}
\end{figure}

In  this subsection, we deal with the simpler case where ${\bf r} = 0$, but $\Delta \eta = \eta - \eta'$ is nonzero. This is the case where the CMB photons
arrive from the same direction, but were emitted at different times and hence different redshifts. In this case there is no angle dependence inside the cosine function in
Eq.~\eqref{eq:K-def}, and we find
 \begin{equation}
 K(k_0\, \Delta \eta,0) = 2\, C_3(k_0\, \Delta \eta)\,,
  \end{equation}
which is plotted in Fig.~\ref{fig:Corr-plot0}.  This function describes the temperature fluctuations correlations for photons arriving from the 
same direction, but different redshifts, with the pre-factor of $1/[a(\eta)\, a(\eta')]$ in Eq.~\eqref{eq:temp-corr2} removed.

Note that the region in which the inflationary gravitons interact with the  CMB photons is actually a last scattering shell of finite thickness. This thickness is determined
by the time required for the electrons and protons in the primordial plasma to combine into neutral hydrogen atoms. It has been estimated by  Hadzhiyska and
 Spergel~\cite{HS19} to be of order $\Delta \eta \approx 19 \,{\rm Mpc} $, which is equivalent to  $\Delta z \approx 90 $ in redshift.

\begin{figure}[htbp]
\includegraphics[scale=0.2]{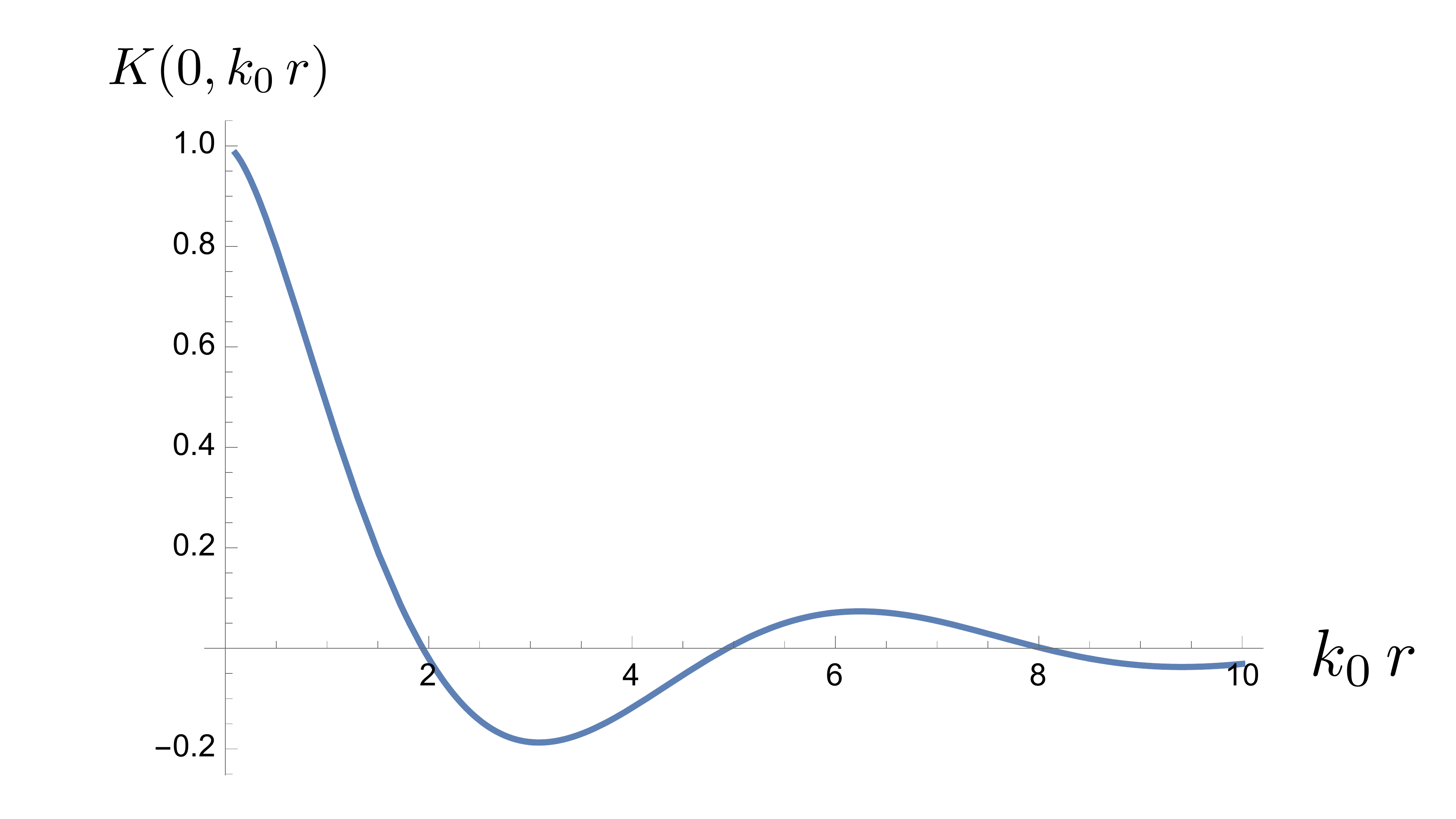}
\caption{ This plot shows the temperature fluctuation correlations between regions at the same redshift, but different spatial locations and seen at different 
angles. Here regions separated by $k_0\, \Delta r \approx 3$ are somewhat anti-correlated.}
\label{fig:Corr-plot1}
\end{figure}

\subsection{Results: Case ${\bf r} \not= 0$ }
  \label{sec:results2}

When ${\bf r} \not= 0$, the argument of the cosine function in  Eq.~\eqref{eq:temp-corr2} depends upon $\theta$.
In this case, we have
 \begin{equation}
 \int_0^{2 \pi}d\phi \int_{-1}^1 dc\,  \sin^4 \chi \, \cos[k(c r - \Delta \eta)] = \frac{4 \pi}{k^5\, r^5} \, \cos(k \Delta \eta) \,[ k r ( 2 k^2 r^2 - 9)\cos(k r)
 +(9 - 5 k^2 r^2  +k^4 r^4) \sin(k r)]\,.
 \end{equation}
 We may now write Eq.~\eqref{eq:K-def} as
  \begin{eqnarray}
 K( \Delta \eta, r) &=& \frac{15}{8}\; \biggl\{   \frac{2}{\rho^2} \, [C_5(\rho+\tau) + C_5(\rho-\tau)] - \frac{9}{\rho^4} \, [C_7(\rho+\tau) + C_7(\rho-\tau)] \nonumber  \\
 &+& \frac{9}{\rho^5} \, [S_8(\rho+\tau) + S_8(\rho-\tau)] - \frac{5}{\rho^3} \,  [S_6(\rho+\tau) + S_6(\rho-\tau)] + \frac{1}{\rho} \,  [S_4(\rho+\tau) + 
 S_4(\rho-\tau)]  \biggr\} \, .
 \label{eq:K}
 \end{eqnarray}
 with $\tau = k_0\, \Delta \eta$ and $\rho = k_0\, r$. 
 
\begin{figure}[htbp]
\includegraphics[scale=0.2]{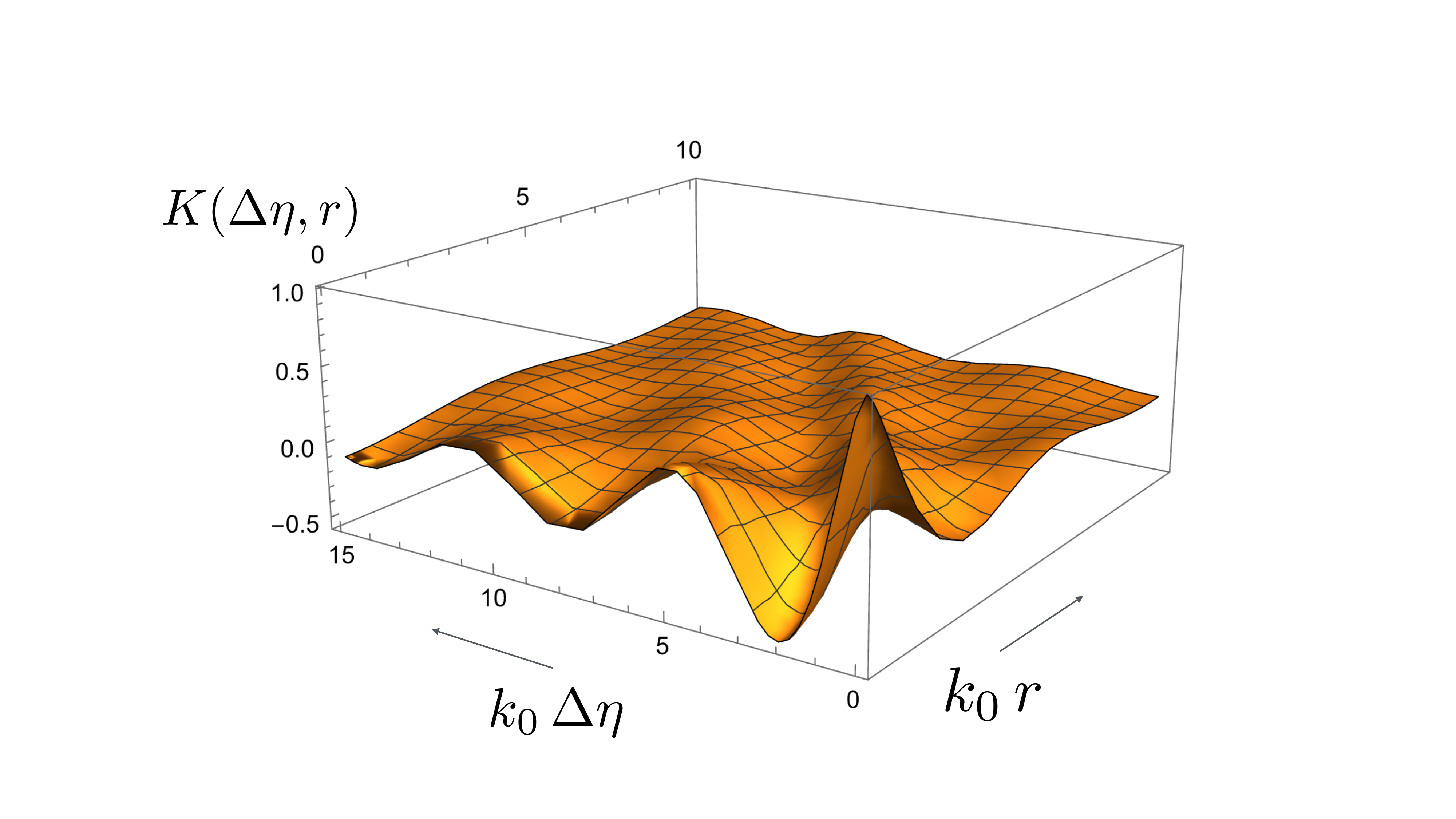}
\caption{ This plot illustrates the relative degrees of correlation and anti-correlation as a function of both spatial and temporal separation. }
\label{fig:Corr-plot2}
\end{figure}

 The case of spatial separation alone, $ K( 0, r) $ is plotted in Fig.~\ref{fig:Corr-plot1}, and the general case is illustrated in  Fig.~\ref{fig:Corr-plot2}.

\subsection{Physical Meaning of the Correlations}
 \label{sec:meaning}

We have seen in the  previous subsection that the temperature fluctuations produced by a graviton bath display distinctive correlations and 
anti-correlations, both in space and in time. The anti-correlations mean that a region whose temperature is above average, is more likely to
be near a region which is cooler than average, and vice versa. The separation between the hotter and the cooler region can be either in angle
or in redshift. The magnitude of the expected angular or redshift depends upon the unknown parameter $k_0$, and is given by $k_0\, \Delta r \approx 3$
for spatial or angular separations and by $k_0\, \Delta \eta \approx 2$ for temporal or redshift separations. Note that the ratio of these two
expected separations, $\Delta r/\Delta \eta \approx 1.5$, is independent of $k_0$. As was emphasized in Sec.~\ref{sec:corr} and illustrated
in Fig.~\ref{fig:corr-flucts}, correlation functions are averages of highly fluctuating quantities. This means that the temperature anti-correlations
can only be observed by averaging over several pairs of regions.

\section{Summary}
\label{sec:final}

We have presented a description of inflationary tensor perturbations in terms of gravitons created by quantum particle creation at the end of inflation.
These gravitons are in a multi-mode squeezed vacuum state describing pairs of created gravitons. However, only one member of each pair is observable,
and the resulting quantum state exhibits large fluctuations and is equivalent to a density matrix .
Our description introduces two infrared cutoffs. The first is the parameter $k_c$ introduced in Sect.~\ref{sec:flat}, and depends upon initial conditions before
the onset of inflation. This parameter is expected to be very small and its effects to be unobservable. The second cutoff is the parameter  $k_0$ introduced in 
Sect.~\ref{sec:k0}, and determined by the range of validity of the concept of graviton number in the out-region. It is an undetermined in our analysis, but is 
expected to be determined by the horizon size at the time of last scattering. We have followed the approach of previous authors, and calculated the CMB temperature
anisotropies from solutions of the photon geodesic equations. This leads to the temperature fluctuation correlation function in terms  of the graviton correlation
function, Eq.~\eqref{eq:grav-corr}. The cutoff $k_0$ is used to make these expressions infrared finite.

An alternative approach may be to use the geodesic deviation equation to study the perturbations of the photon geodesics. This would replace the correlation function
for the graviton field by that for the associated Riemann tensor. The letter contains additional derivatives which are expected to render it infrared finite. This approach will be
the topic for future research.

Within our present approach, we are able to obtain an explicit expression, Eq.~\eqref{eq:K}, as a function of the temporal separation $ \Delta \eta$ and spatial separation $r$
of a pair of emission points. Recall that $r$ is equivalent to an angular separation on the sky, and $ \Delta \eta$ is equivalent to a redshift separtion within the last scattering
shell. The plots of this function in Figs.~\ref{fig:Corr-plot0}, \ref{fig:Corr-plot1}, and \ref{fig:Corr-plot2}, reveal significant anti-correlations in both angle and redshift. This means
that a hot region will tend to be associated with a nearby cooler region, and vice versa. As was emphasized in Sect.~\ref{sec:corr}, these anti-correlations will appear only in averages 
over many pairs of emission points. However, if the  anti-correlations can be observed, it will provide strong evidence for the existence of inflationary gravitons.

\begin{acknowledgments} 
We would like to thank J.T. Hsiang, B.L. Hu, and K.W. Ng for helpful discussions. LHF would like to thank
the Department of Physics, Soochow University for hospitality while part of this work was performed.
This work was supported in part  by the National Science Foundation under Grant PHY-2114024.
\end{acknowledgments}

 \appendix
 \section{Integrals Appearing in the Correlation Functions}

Here we give the explicit forms for some of the integrals which appear in the correlation function, Eq.~\eqref{eq:K}.
For $n$ odd, $C_n(\alpha)$, defined in Eq.~\eqref{eq:Cn}, is given by (See Formula 3.761.8 in Ref,~\cite{GR}.)
\begin{equation}
 C_{2m+1}(\alpha) = \frac{\alpha^{2 m}}{(2 m)!} \, \left\{ \sum_{k=1}^{2m} \frac{(2m-k)!}{\alpha^{2m-k+1}} \, \cos[\alpha +(k-1) \frac{\pi}{2} ] + (-1)^{m+1} \, {\rm ci}(\alpha) \right\}\,,
 \end{equation}
  where ${\rm ci}(\alpha) = - \int_\alpha^\infty dt \, \cos (t)/t$ is the cosine integral function. The special cases of interest to us are the following:
  \begin{equation}
 C_3(\alpha) = \frac{1}{2} \, \left[ \cos(\alpha) - \alpha\, \sin(\alpha) + \alpha^2 \, {\rm ci}(\alpha)  \right]\,,
 \end{equation} 
   \begin{equation}
 C_5(\alpha) = \frac{1}{4!} \, \left[  3!\, \cos(\alpha) - 2 \alpha \,  \sin(\alpha)  - \alpha^2 \, \cos(\alpha) + \alpha^3 \,  \sin(\alpha)  - \alpha^4 \, {\rm ci}(\alpha)  \right]\,,
 \end{equation}
  and
   \begin{equation}
 C_7(\alpha) = \frac{1}{6!} \, \left[  5!\, \cos(\alpha) - 4!  \alpha \,  \sin(\alpha)  -  3!\, \alpha^2 \, \cos(\alpha) + 2 \,\alpha^3 \,  \sin(\alpha)  +  \alpha^4 \, \cos(\alpha) 
   - \alpha^5 \, \sin(\alpha) +\alpha^6\,   {\rm ci}(\alpha)  \right]\,.
 \end{equation}
 
 For $n$ even, $S_n(\alpha)$, defined in Eq.~\eqref{eq:Sn},  is given by (See Formula 3.761.3 in Ref,~\cite{GR}.)
  \begin{equation}
 S_{2m1}(\alpha) = \frac{\alpha^{2 m-1}}{(2 m-1)!} \, \left\{ \sum_{k=1}^{2m-1} \frac{(2m-k)!}{\alpha^{2m-k}} \, \sin[\alpha +(k-1) \frac{\pi}{2} ] + (-1)^{m} \, {\rm ci}(\alpha) \right\}\,,
 \end{equation}
 which leads to
 \begin{equation}
 S_4(\alpha) = \frac{1}{6} \, \left[ 2\, \sin(\alpha)  + \alpha\, \cos(\alpha) - \alpha^2\, \sin(\alpha) + \alpha^3 \, {\rm ci}(\alpha)  \right]\,,
 \end{equation}  
 \begin{equation}
 S_6(\alpha) = \frac{1}{5!} \, \left[ 4!\, \sin(\alpha)  +  3!\, \alpha\, \cos(\alpha) -  2 \, \alpha^2\, \sin(\alpha)   -  \alpha^3\, \cos(\alpha)    + \alpha^4\, \sin(\alpha)    - \alpha^5 \, {\rm ci}(\alpha)  \right]\,,
 \end{equation}  
 and
 \begin{equation}
 S_8(\alpha) = \frac{1}{7!} \, \left[ 6!\, \sin(\alpha)  +  5!\, \alpha\, \cos(\alpha) -  4! \, \alpha^2\, \sin(\alpha)   - 3!\, \alpha^3\, \cos(\alpha)    + 2\, \alpha^4\, \sin(\alpha)  +\alpha^5\,  \cos(\alpha) - \alpha^6\, \sin(\alpha) + \alpha^7 \, {\rm ci}(\alpha)  \right]\,.
 \end{equation}

\end{document}